# Learning Power Control from a Fixed Batch of Data

Mohammad G. Khoshkholgh, *Member, IEEE,* and Halim Yanikomeroglu, *Fellow, IEEE*

*Abstract*—We address how to exploit power control data, gathered from a monitored environment, for performing power control in an unexplored environment. We adopt offline deep reinforcement learning, whereby the agent learns the policy to produce the transmission powers solely by using the data. Experiments demonstrate that despite discrepancies between the monitored and unexplored environments, the agent successfully learns the power control very quickly, even if the objective functions in the monitored and unexplored environments are dissimilar. About one third of the collected data is sufficient to be of high-quality and the rest can be from any sub-optimal algorithm.

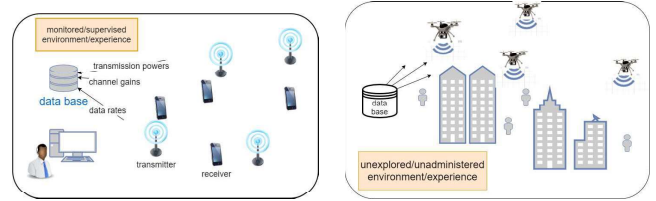

Fig. 1. The monitored environment (LHS), could be a simulation setup too, is abstracted via a data set comprising of hundreds of thousands of records, each of which representing (at least) the channel power gains, the transmission power of the users, and the sum capacity. The data set is used to learn power control in the unexplored environment (RHS).

## I. INTRODUCTION

DEEP reinforcement learning (DRL) has demonstrated tremendous success in solving very intricate decision making problems in gaming, computer vision, robotics, health care, and finance [1]. Recent developments have also highlighted various potentials of DRL for resource allocation in communication networks [2]. DRL-based approach to the network managements departs from the conventional ones that are based on solving well-crafted optimization problems—heavily depending on the availability of tractable mathematical models of the wireless networks.

Among one of the complex resource allocation problems in communication networks is power allocation in interference channel [3]. The optimization problem in its most basic form, which is the maximization of the sum rate subject to the maximum permissible transmission power of the transmitters, is hard to solve due to its non-convexity. When the channel state information (CSI) is not known perfectly the conventional solutions become almost intractable, hence methods based on DRL seems promising. The use of deep Q-learning (DQN) to derive near optimal power allocation in the cellular network is discussed in [4]–[6]. It is shown that DQN is effective to deal with complexity of power allocation in multi-cell downlink cellular networks. However, DQN requires to firstly quantized the transmission powers into a fixed number of bins. As the transmission power is a continuous variable, its discretization could be arbitrary and may lead to curse of dimensionality [7]. The use of continuous DRL for the continuous power control in interference channel is discussed in [8]. On the other hand, in practice the application of such algorithms seem limited due chiefly to sample inefficiency. The agent goes under try and fail steps for up to millions of iterations before reaching a convergence point. In effect, conventional DRL algorithms are designed to deal with the problem, e.g., power control, from scratch, ignoring existing engineering insights/knowledge that can be harnessed from a *supervised/monitored environment/experience*[1] in the form of data. Our main goal here is to demonstrate how the pervious data can be used to (quickly) learn the continuous power control in an *unadministered/unexlored environment/experience* without further interaction of the agent with the environment, see Fig. 1. For this goal, we adopt *offline*[2] DRL, whereby the agent learns the policy to produce the actions (transmission powers) solely based on the data of the monitored environment. Particularly, we use Batch Constrained Q-learning (BCQ) algorithm [9], which is mainly developed for the locomotion tasks. Our experiments demonstrate that despite discrepancies between the monitored and unexplored environments, the agent successfully learns the power control very quickly.

## II. SYSTEM MODEL

We focus on the interference channel power control problem consisting of $K$ single-antenna transceivers. Each transmitter $k$ has its own intended receiver with $h_{kk} \in \mathbb{R}^+$ standing as the corresponding channel power gain. Transmitter $k$ poses interference on the other receivers through channel power gains $h_{kj} \in \mathbb{R}^+$. We assume the interference is considered as noise. The experienced signal-to-interference-plus-noise (SINR) at the receiver $k$ is $\text{SINR}_k = \frac{h_{kk}P_k}{\sigma_k^2 + \sum_{j \neq k} h_{jk}P_j}$ where $\sigma_k^2$ is the noise power at the receiver $k$ and $P_k \in [0, \hat{P}_k]$ is the (continuous) transmission power, which should be smaller than the maximum permissible transmission power $\hat{P}_k$. The data rate of user $k$ is calculated by the Shannon's formula $r_k = \log(1 + \text{SINR}_k)$. Assuming that CSI is perfectly known at the transmitters as well as the receivers, the optimal

The authors are with the Department of Systems and Computer Engineering, Carleton University, Ottawa, Ontario, Canada. Email: m.g.khoshkholgh@gamil.com, halim @sce.carleton.ca.

This work was supported by Huawei Canada Co., Ltd.

[1] In this paper, experience and environment has the same implication.
[2] Offline DRL differs from the off-policy policy gradient such as DDPG [7], whereby the agent is allowed to collect more data while using a experience buffer for learning.

2                                                                                                                    IEEE WIRELESS COMMUNICATIONS LETTERS

power allocation can be obtained by solving the following optimization problem:

$$\max_{P_1,\ldots,P_K} \sum_k \log(1 + \text{SINR}_k) \quad \text{s.t.} \quad 0 \leq P_k \leq \hat{P}_k, \quad \forall k.$$

We denote this environment by $\mathcal{E}_M$, which can be a representation of a simulation setup, a previous experience, or a monitored environment. Running a power control algorithm, e.g., WMMSE [3], a data set $\mathcal{D} = \{s_t, a_t, r_t, s_{t+1}\}_t$ with size of $|\mathcal{D}|$ is collected, where $s_t$ is the state at time slot $t$, containing all the channel gains, and $a_t$ is the transmission powers. The sum capacity is considered as reward $r_t$. The choice of the power control algorithm in the controlled environment is not relevant to the agent. The agent desires achieving a similar performance that the data represents without making any assumption regarding the way that the data is collected. It is basically up to the administrator of the controlled environment to ensure the quality of the data.

## III. BACKGROUND

An agent operating in an uncertain environment when the state and action spaces are continuous is the focus of continuous DRL [7]. Upon observing state $s_t \in \mathbb{R}^S$ ($S$ is the dimension of the state space) at time slot $t$, the agent takes an action $a_t \in \mathbb{R}^B$ ($B$ is the action dimension), which leads to the new state $s_{t+1}$ and the bounded reward $r_t(s_t, a_t) \in \mathbb{R}$, or simply $r_t$. The ultimate goal is to maximize the expected aggregate discounted reward $J(\pi) = \mathbb{E}_\pi \sum_t \gamma^t r_t(s_t, a_t)$, also known as return [1], by finding an optimal policy $\pi_\phi(s_t)$. Parameter $\gamma \in (0, 1]$ is the discount factor prioritizing short-term rewards. The policy guides the agent to what action should be taken in a certain state in order to maximize the reward. Policy $\pi_\phi$ is approximated via high-capacity non-leaner function approximator such as deep neural network (DNN) [10] with parameter $\phi$. The policy has its value function, also known as Q-function, $Q^\pi(s_t, a_t) = \mathbb{E} \sum_{t' \geq t} \gamma^{t'-t} r_{t'}(s_{t'}, a_{t'})$, which measures the return by taking action $a_t$ when the agent is in state $s_t$. As the state/action spaces are (large) continuous, the value function is estimated via DNN with parameters $\theta$, denoted by $Q_\theta$. The goal is then to learn the DNNs via series of interactions with the environment. Commonly, the interactions are recorded in a replay buffer $\mathcal{B}$ comprising of tuples $(s_t, a_t, r_t, s_{t+1})$. Sampling the replay buffer in each iteration the value function $Q_\theta$ is then updated by calculating the target value [1], [2]:

$$y_t = r_t + \gamma Q_{\theta'}(s_{t+1}, \pi_\phi(s_{t+1})). \tag{1}$$

Target and the current value function $Q_\theta(s_t, a_t)$ are used to update the Q-function via mean-square-error regression [7]. Here, to stabilize the algorithm the use of the target network with parameters $\theta'$ is considered. Note that the target network does not need to be learned, as its parameters are updated via $\theta' \leftarrow \tau\theta + (1-\tau)\theta'$ in each iteration, where $\tau$ is a small parameter close to 0.005 [2]. Finally, the policy parameters are updated via gradient ascent $\phi \leftarrow \arg\max_\phi \sum Q_\theta(s_t, \pi_\phi(s_t))$. Note that under the conventional DRL setup the size of the replay buffer grows during the course of learning, and when it reaches its maximum size the old records are replaced with the newly collected ones. In short, the agent is free to interact with the environment, which is not the case of offline learning.

In the remainder of this paper, to avoid overloading the notations, we drop the time subscript $t$ and use $s$ instead of $s_t$. Furthermore, we denote $s'$ as the next state.

## IV. POWER CONTROL FROM DATA

Our goal is to use the data $\mathcal{D}$ gathered from $\mathcal{E}_M$ (see Fig. 1). We here adopt BCQ algorithm, which is one of the state-of-the-art offline DRL algorithms developed for robotics applications (continuous action space) [9]. Under BCQ, the agent must learn four DNNs: an actor parameterized by $\phi$, a generative model parameterized by $\omega$, and finally two critics or value networks with parameters $\theta_1$ and $\theta_2$. The learning is confined to the use of the database $\mathcal{D}$; no new interactions with the unexplored environment is conducted. The pseudo-code of the algorithm can be found in Algorithm 1.

*1) Generative Network:* If the agent adopts a typical off-policy algorithm, such as DDPG [7], the performance becomes very poor as the true policy that renders the maximum reward in the new environment becomes uncorrelated with the one that can be extracted from the data. This is due to the fact that the state-action pairs which selected by the agent can diverge substantially from the ones existing in the data, leading to very poor estimation of $Q_\theta(s', \pi_\theta(s'))$. One way to account for this issue is via choosing actions that are more probable to happen for a given state, via maximizing the marginal likelihood $a \leftarrow \arg\max_a P(a|s)$, where $P(.|.)$ is the conditional probability distribution. Nevertheless, as the state/action spaces are continuous this maximization is intractable, a parametric generative model $G_\omega(s)$, parameterized by $\omega$, is used instead. Here, the generative model is represented via conditional variational auto-encoder (CVAE), which, in essence, presents the conditional distribution $P(a|s)$ via transforming the underlying latent[3] variable $z$ [11]. Thus, rather than directly coming up with an expression for $P(a|s)$, the CVAE models a computationally affordable joint distribution $P(a|s) = \int P(a|s,z)P(z|s)dz$, where $P(z|s)$ is the conditional priori which is modelled as a standard Gaussian distribution. The generative distribution $P(a|s,z)$ governs the production of actions for the given state and latent variable. We should compute the generative distribution through the posteriori $P(z|s,a)$ and a priori. However, the posteriori is also intractable and then is modelled via the recognition/encoder $E_{\omega_e}$, which is a DNN with parameters $\omega_e$. This implies that the actual posteriori $P(z|s,a)$ is approximated via $q_{\omega_e}(z|s,a)$. Thus: $(\mu, \sigma) = E_{\omega_e}(s, a)$ and $q_{\omega_e}(z|s,a) = \mathcal{N}(\mu, \sigma^2)$. The generative model, known as decoder, is also modelled via a DNN with parameters $\omega_d$. To efficiently learn the parameters of the encoder and decoder, the maximization of the variational lower bound is used [11]:

$$\log P(a|s) \geq -D_{KL}(q_{\omega_d}(z|s,a)||P_{\omega_e}(z|s)) \tag{2}$$
$$+ \mathbb{E}_{q_{\omega_d}(z|s,a)} \log P_{\omega_e}(a|s,z),$$

---

[3]Latent variables are not part of the data set so we do not observe them. We consider them as part of the model.

where the first term measures the distance between distributions $q_{\boldsymbol{\omega}_d}(\boldsymbol{z}|\boldsymbol{s},\boldsymbol{a})$ and $P_{\boldsymbol{\omega}_e}(\boldsymbol{z}|\boldsymbol{s})$, which is known as KL divergence[4]. Under the Gaussian distribution assumption, the first term can be evaluated in a closed form as a function of mean and standard deviation of the underlaying Gaussian distributions. However, the second term should be evaluated via Monte Carlo sampling from the recognition network $q_{\boldsymbol{\omega}_d}(\boldsymbol{z}|\boldsymbol{s},\boldsymbol{a})$. It is, nevertheless, technically problematic to draw samples from this posteriori. Instead, initial samples $\boldsymbol{y}^l$, considered as noise, are drawn from a standard Gaussian distribution. These samples are then transformed via a deterministic, differentiable function $g_{\boldsymbol{\omega}_d}(\boldsymbol{a},\boldsymbol{s},\boldsymbol{y}^l)$, which is indeed the function that is approximated via DNN. This transformation is essential for the back-propagation of the error through the Gaussian latent variables. Consequently, an empirical variational lower bound is obtained from

$$\log P(\boldsymbol{a}|\boldsymbol{s}) \gtrapprox -D_{KL}(q_{\boldsymbol{\omega}_d}(\boldsymbol{z}|\boldsymbol{s},\boldsymbol{a})||P_{\boldsymbol{\omega}_e}(\boldsymbol{z}|\boldsymbol{s})) \quad (3)$$

$$+\frac{1}{L}\sum_{l=1}^{L}\log P_{\boldsymbol{\omega}_e}(\boldsymbol{a}|\boldsymbol{s},\boldsymbol{z}^l), \ \boldsymbol{z}^l = g_{\boldsymbol{\omega}_d}(\boldsymbol{a},\boldsymbol{s},\boldsymbol{y}^l), \boldsymbol{y}^l \sim \mathcal{N}(\boldsymbol{0},\boldsymbol{I}),$$

where $L$ is the number of samples. The loss function that is used for training the generative model is[5] $\boldsymbol{\omega} \leftarrow \arg\min_{\boldsymbol{\omega}} \sum(\tilde{\boldsymbol{a}}-\boldsymbol{a})^2$, see line 7 of Algorithm 1.

The training of the generative model empowers the agent to produce relevant actions based on the received state, which can resemble the state-actions pairs that can be found in the data set. In details, for the state $\boldsymbol{s}$, from the conditional priori $P_{\boldsymbol{\omega}_e}(\boldsymbol{z}|\boldsymbol{s})$ the latent random variable $\boldsymbol{z}$ is produced. The action $\boldsymbol{a}$ is then drawn from the generative distribution $P_{\boldsymbol{\omega}_e}(\boldsymbol{a}|\boldsymbol{s},\boldsymbol{z})$. The set of candidate actions for the given state is then used by the DQN to train the actor and critic networks.

*2) Actor:* The generative model $G_{\boldsymbol{\omega}}(\boldsymbol{s})$ along with the critic model $Q_{\boldsymbol{\theta}_1}(\boldsymbol{s},\boldsymbol{a})$ is used as part of the policy to derive $n$ plausible actions by sampling the generative model $n$ times. For each drawn sample the associated value function should be evaluated and the best action that renders the maximum value function $Q_{\boldsymbol{\theta}_1}(\boldsymbol{s},\boldsymbol{a})$ is chosen as the action. Still there is the possibility of poor training performance due to lack of diversity in candidate actions. As a remedy, the candidate actions are perturbed by a DNN $\zeta_{\boldsymbol{\phi}}(\boldsymbol{s},\boldsymbol{a},\Phi)$, where $\Phi$ is a hyper-parameter controls the level of the perturbation affected the actions. Note that the output of the perturbation network stay in range $[-\Phi,\Phi]$. Effectively, the policy in BCQ is governed by

$$\pi_{\boldsymbol{\theta}} = \arg\max_{\boldsymbol{a}'_i}Q_{\boldsymbol{\theta}_1}(\boldsymbol{s},\boldsymbol{a}'_i) \quad (4)$$
$$\{\boldsymbol{a}'_i = \boldsymbol{a}_i + \zeta_{\boldsymbol{\phi}}(\boldsymbol{s},\boldsymbol{a}_i,\Phi) : \boldsymbol{a}_i \sim G_{\boldsymbol{\omega}}(\boldsymbol{s})\}_{i=1}^n. \quad (5)$$

A recommended choice of $n$ is 10 [9].

---
[4]For probability distributions $P$ and $Q$ over a given random variable the KL divergence is defined as $D_{KL}(P||Q) = \mathbb{E}_P[\log\frac{P}{Q}]$.

[5]In the original BCQ algorithm, the loss function contains an extra term related to the KL divergence of the distributions $\mathcal{N}(\mu,\sigma)$ and $\mathcal{N}(0,1)$. Our experiments show that the inclusion of this extra loss function causes poor performance, so we remove it.

*3) Critics:* In BCQ, two critic networks $Q_{\boldsymbol{\theta}_1}(\boldsymbol{s},\boldsymbol{a})$ and $Q_{\boldsymbol{\theta}_2}(\boldsymbol{s},\boldsymbol{a})$ should be learned. However, compared to conventional Q-learning algorithms that the target value $y$ is estimated via Eq. 1, in BCQ target value is calculated as

$$y_t = r_t + \gamma \max_{\boldsymbol{a}'_i}(\lambda \min_j Q_{\boldsymbol{\theta}'_j}(\boldsymbol{s}',\boldsymbol{a}'_i) + (1-\lambda)\max_j Q_{\boldsymbol{\theta}'_j}(\boldsymbol{s}',\boldsymbol{a}'_i).$$

For given value of $\lambda \in [0,1]$, the inner term represents a convex combination of minimum and maximum of target critics. The minimum of Q values are included to compensate for the possible bias in the estimation of the value functions. On the other hand, it helps to penalize the high variance induced by poor actions that are not tended to be found in the data set. Nevertheless, to not overcompensate for the future uncertainties in estimating the Q values the parameter $\lambda$ is chosen around 0.75.

## V. EXPERIMENTS

For the experiments we use the pytorch library [12]. For each experiment we consider 10 different random seeds and calculate the average values accordingly.

*BCQ Networks:* Policy is modelled via DNN with 3 dense layers with input/output dimensions $(S+B)/400$, $400/300$, and $300/B$, respectively, where $S = K^2$ is the space dimension and $B = K$ is the action dimension. Similarly, the critics are modelled as DNNs with three dense layers $(S+B)/400$, $400/300$, and $300/1$, respectively. The activation functions are ReLU [10]. The encoder is a DNN with three dense layers. The first and second layers are with input/output dimensions $(S+B)/750$ and $750/750$, respectively. This DNN has two heads, one for the mean value and the other for the logarithm of the standard deviation. Each of these are modelled by its associated dense layer with size $750/C$, where $C = 2B$ is the dimension of the latent space. The decoder is also a DNN with three dense layers with dimensions $(C+S)/750$, $750/750$, and $750/B$, respectively.

*Monitored Environment $\mathcal{E}_M$:* The monitored environment contains the previous experiment in which the WMMSE [3] is adopted to obtain the transmission powers. The pertinent measurements including states, actions, and rewards are collected

---

**Algorithm 1** BCQ (slightly modified)

1: Hyper-parameters: Data set size $|\mathcal{D}|$, horizon $T$, target network update rate $\tau$, mini-batch size $N$, max perturbation $\Phi$, number of sampled actions $n$, minimum weighting $\lambda$
2: Input: initialize Q-networks $Q_{\boldsymbol{\theta}_1}$, $Q_{\boldsymbol{\theta}_1}$, perturbation network $\zeta_{\boldsymbol{\phi}}$, and VAE $G_{\boldsymbol{\omega}} = \{E_{\boldsymbol{\omega}_e}, D_{\boldsymbol{\omega}_d}\}$ with random parameters $\boldsymbol{\theta}_1$, $\boldsymbol{\theta}_2$, $\boldsymbol{\phi}$, and $\{\boldsymbol{\omega}_e,\boldsymbol{\omega}_d\}$
3: Set target network parameters $\boldsymbol{\theta}'_1 \leftarrow \boldsymbol{\theta}_1$, $\boldsymbol{\theta}'_2 \leftarrow \boldsymbol{\theta}_2$, and $\boldsymbol{\phi}' \leftarrow \boldsymbol{\phi}$
4: **for** $t = 1,2,\ldots T$ **do**
5:    Sample mini-batch $\mathcal{B} = \{\boldsymbol{s},\boldsymbol{a},r,\boldsymbol{s}'\}$ from $\mathcal{D}$ randomly
6:    $\mu,\sigma = E_{\boldsymbol{\omega}_e}(\boldsymbol{s},\boldsymbol{a})$, $\tilde{\boldsymbol{a}} = D_{\boldsymbol{\omega}_d}(\boldsymbol{s},\boldsymbol{z})$, $\boldsymbol{z} \sim \mathcal{N}(\mu,\sigma)$
7:    $\boldsymbol{\omega} \leftarrow \arg\min_{\boldsymbol{\omega}}\sum(\tilde{\boldsymbol{a}}-\boldsymbol{a})^2$
8:    Sample n actions $\{\boldsymbol{a}_i = G_{\boldsymbol{\omega}}(\boldsymbol{s}')\}_{i=1}^n$
9:    Perturb each action $\{\boldsymbol{a}_i = \boldsymbol{a}_i + \zeta_{\boldsymbol{\phi}}(\boldsymbol{s}',\boldsymbol{a}_i,\Phi)\}_{i=1}^n$
10:   Compute Q-value $q = \max_{\boldsymbol{a}_i}(\lambda\min_j Q_{\boldsymbol{\theta}'_j}(\boldsymbol{s}',\boldsymbol{a}_i) + (1-\lambda)\max_j Q_{\boldsymbol{\theta}'_j}(\boldsymbol{s}',\boldsymbol{a}_i)$
11:   Compute target $y = r + \gamma q$
12:   $\{\boldsymbol{\theta}_1,\boldsymbol{\theta}_2\} \leftarrow \arg\min_{\{\boldsymbol{\theta}_1,\boldsymbol{\theta}_2\}}\sum\sum_{j=1,2}(y-Q_{\boldsymbol{\theta}_j}(\boldsymbol{s},\boldsymbol{a}_i))^2$
13:   $\boldsymbol{\phi} \leftarrow \arg\max_{\boldsymbol{\phi}}\sum Q_{\boldsymbol{\theta}_1}(\boldsymbol{s},\boldsymbol{a}_i + \zeta_{\boldsymbol{\phi}}(\boldsymbol{s},\boldsymbol{a}_i,\Phi))$, $\boldsymbol{a}_i = G_{\boldsymbol{\omega}}(\boldsymbol{s})$
14: **end for**

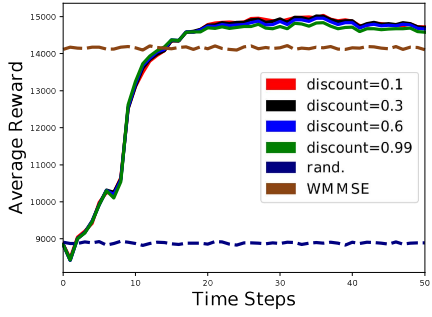

Fig. 2. Average reward over iterations for different values of discount factor, $\gamma$.

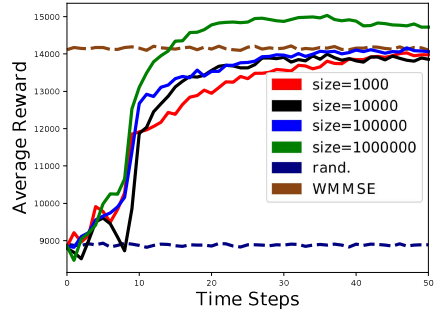

Fig. 3. Average reward over iterations for different values of database size.

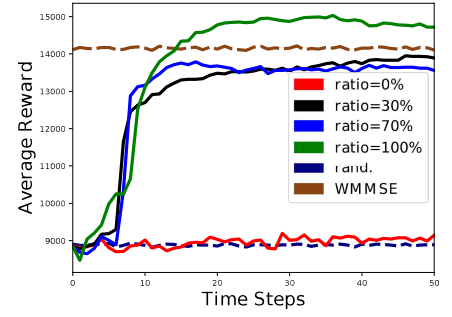

Fig. 4. Average reward over iterations for different types of the data quality.

in the data set $\mathcal{D}$. We consider a circular area with radius 60 m, and randomly locate 4 transmitters and 4 receivers in it. For training we set $\hat{P}_k = 1$ W, $\forall k$. The wireless channel is based on 3GPP Line-of-Sight (LOS) / none-LOS (NLOS) path-loss attenuation model $L_{jk} = \|X_{jk}\|^{-\alpha_l} \sim p_l(\|X_{jk}\|)$ for $l \in \{L, N\}$ [13], where $p_\mathrm{N}(\|X_{jk}\|) = 1 - p_\mathrm{L}(\|X_{jk}\|)$ is the probability of LOS that is a function of distance $\|X_{jk}\|$: $p_\mathrm{L}(\|X_{jk}\|) = \min\left\{\frac{D_0}{\|X_{jk}\|}, 1\right\}\left(1 - e^{-\frac{\|X_{jk}\|}{D_1}}\right) + e^{-\frac{\|X_{jk}\|}{D_1}}$, which is also known as ITU-R UMi model. Also, $\alpha_\mathrm{L}$ (resp. $\alpha_\mathrm{N}$) is the path-loss exponent associated with LOS (resp. NLOS) component where $\alpha_\mathrm{N} > \alpha_\mathrm{L}$. $D_1$ and $D_0$ are hyperparameters which can take different values for different environments. We set the channel parameters as $\alpha_L = 2.4$, $\alpha_N = 3.78$, $D_0 = 18$ m, $D_1 = 36$ m, and the background noise power $-173$ dBm/Hz. The fading power gain under the LOS mode is modelled by Nakagami-m distribution with parameter $m = 10$. Under the NLOS mode the fading is modelled via unit-mean exponential random variable. We also consider large-scale shadowing with mean zero dB and standard deviation 5 dB under LOS mode and 8.6 dB under NLOS mode. Receivers and transmitters are allowed to dislocate by up to 5 meters in a random direction at the start of each iteration. However, we are making sure that the receivers stay in the simulation area.

*Hyper-Parameters:* The main hyper-parameters we are interested to discuss their impacts on the performance of BCQ are 1) the discount factor $\gamma$, 2) the size of the data set $\mathcal{D}$, and 3) the quality of the collected data. Here we assume that the unexplored environment $\mathcal{E}_U$ is the same as the monitored environment $\mathcal{E}_M$.

In Fig. 2 we study the impact of $\gamma$ on the convergence of BCQ. We consider three values for $\gamma$. As seen, for all the considered values the convergence and the performance is almost the same. On the other hand, for the four first steps, BCQ performs as the random algorithm, by which the transmitters randomly choose their transmission powers. However, the performance quickly peaks up and by the 14th step it reaches the performance of WMMSE. Note that for the convergence of WMMSE we require up to 100 iterations in our experiments, therefore BCQ is able to converge much faster than WMMSE. Interestingly, BCQ improves the reward by about 8% compared to the WMMSE, without resorting to further interactions with the environment.

Now we study the impact of the database size on the performance of BCQ in Fig. 3. Here we set $\gamma = 0.1$. As seen, by reducing the database size the performance of BCQ reduces marginally. On the one hand, it takes a little bit longer for the algorithm to converge–from 14 in the case of $\mathcal{D} = 10^6$ to around 40 in the case of $\mathcal{D} = 10^3$. This is understandable, as the BCQ is in core relies on regression to update the DNNs. On the other hand, when we reduce the size of the database, the performance of BCQ reaches at most to the performance of WMMSE, which is 8% reduction compared to the case of $\mathcal{D} = 10^6$. This experiment implies that the size of the database is not too crucial provided that it is larger than a minimum size. This has important implications for the applications that the storage size of the device are limited or when the collection of data is expensive.

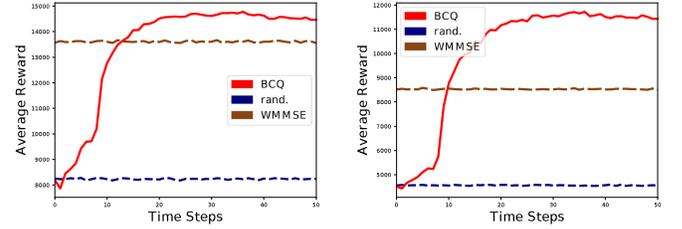

Fig. 5. (a): Average reward of short packet communication system. (b): average reward of UAV-assisted communication.

Finally, we study the quality of the data set on the performance of BCQ. We set $\gamma = 0.1$ and $\mathcal{D} = 10^6$. But instead of filling the data set only from the WMMSE algorithm, we include a percentage of the collected data from the random power allocation algorithm too. In effect, we consider 4 different scenarios: 1) 100% WMMSE data, 2) 70% WMMSE data, 3) 30% WMMSE data, 4) 0% WMMSE data (random power allocation). As seen, by reducing the contribution of the WMMSE algorithm in the collected data, the performance of BCQ reduces. Surprisingly, even only 30% WMMSE data is sufficient to achieve almost 95% of the WMMSE performance. It implies that even very poor collection of data is still valuable for achieving good performance. However, for the case of 0% WMMSE data, BCQ could not do better than the random power allocation. The findings of this experiment has crucial implications, noting that in practice the data could be gathered from imperfect monitored experience. Such databases are still valuable for learning.

*Different Objective Function:* In the monitored environment the data set was filled with the power allocation vector optimizing the Shannon's capacity, which targets the communication paradigm with relatively long transmission delay and large packet size. In small packet communication applications, the Shannon capacity formula is no longer applicable, and decoding error probability cannot be vanishingly small. The short packets are usually subject to extremely low-latency transmission delay, for instance, for control information delivery, known as ultra reliable low latency communication (URLLC). Here, we are interested to see whether the collected data has any value for short packet communications.

Assume that the packet length is M=200 bits and the required decoding error is $\epsilon = 10^{-9}$. According to [14] the achievable data rate can be obtained from $r_k = \log(1 + \text{SINR}_k) - \frac{1}{\ln 2}\sqrt{\frac{V_k}{M}}Q^{-1}(\epsilon)$, where $Q^{-1}(x)$ is the inverse of $Q(x) = \frac{1}{\sqrt{2\pi}}\int_x^\infty e^{-0.5t^2}dt$, and $V_k$ is the channel dispersion and is $V_k = 1 - (1 + \text{SINR}_k)^{-2}$. As seen, compared to the Shannon's capacity the achievable data rate is subject to a penalty term associated with the short packet length.

In Fig. 5-(a) we show the results. As seen, the BCQ algorithm is able to utilize the data to learn the power allocation for new objective function. On the other hand, the performance is higher compared to the WMMSE algorithm, which is not optimal for this case. As a result, BCQ is able to successfully cope with changed reward function, which enhances the usability of the previous data for much broader applications.

*Different Environment:* Now we explore the advantage of BCQ for extending the usage of the data for different environments. Here, we consider the case that the transmitters are UAV. Let $H_j \in [40, 120]$ be the altitude (in meters) of the UAV $j$, which is randomly selected at each time step. All UAVs are equipped with a directional antenna with a circular radiation pattern and beam-width of $\omega = \pi/3$. We denote the main-lobe and side-lobe antenna gain for UAV $j$ by $G$ and $g \neq 0$, respectively, where $g = G/40$ and $G = 2.6/\omega^2$. The vertical angel between the receiver $k$ and the UAV $j$ is $\rho_{jk} = \tan^{-1}(H_j/\|X_{jk}\|)$. The receiver $k$ is within the main-lobe of UAV $j$, if $\rho_{jk} > \pi/2 - \omega/2$, or equivalently for $\|X_{jk}\| < \frac{H_j}{\tan(\pi/2-\omega/2)}$ [15]. The probability that the channel between UAV $j$ and the receiver is in LOS status is obtained from $p_L(\|X_{jk}\|) = \left(1 + \phi e^{-\psi\left(\frac{180}{\pi}\arctan(\frac{H_j}{\|X_{jk}\|})-\phi\right)}\right)^{-1}$ [16], where $\phi$ and $\psi$ are the channel parameters representing the characteristics of the communication environment. We set $\phi = 27.23$ and $\psi = 0.08$ which corresponds to high-rise environment [16]. Note that the 3-D distance between UAV $j$ and user $k$ is $\sqrt{H_j^2 + \|X_{jk}\|^2}$. Fast fading stays as the case of monitored environment. The log-normal gain is also modelled via $V_{jk} = 10^{U_{jk}/10}$ where $U_{jk} \sim \mathcal{N}(\mu^l, \sigma_{jk}^l)$ in which $\sigma_{jk}^l = a_l e^{-c_l \frac{180}{\pi}\arctan(\frac{H_j}{\|X_{jk}\|})}$ [17]. We set $\mu_L = 1.5$, $\mu_N = 29$, $a_L = 7.37$, $a_N = 37.08$, $c_L = 0.03$, and $c_N = 0.03$, which correspond to the high-rise environment [16].

Fig. 5-(b) shows the results. As seen, the agent is able to learn the power allocation very quickly, despite the fact that the unexplored environment is entirely different than the monitored environment. On the other hand, we note that under WMMSE the reward is much lower than that of BCQ that highlights the power of BCQ algorithm.

## VI. CONCLUSIONS

We adopted offline DRL to learn the power control from a fixed batch of data without further interaction with the environment. Our simulations showed that the learning speed is very fast and is robust against the quality of the data: Only 30% of the data needs to be collected from WMMSE for achieving 95% of the WMMSE's performance. On the other hand, database with as small as 1000 records seemed to be sufficient to achieve about 90% of the optimal performance. We also observed that the reward of the agent can be different than the reward that the data was collected based on, leading to exploiting data more efficiently practically.